\documentclass[showpacs,twocolumn,prl,aps]{revtex4}
\usepackage{graphicx}

\begin{document}

\title[Short title for running header]{Chiral d-wave RVB state on honeycomb lattice as a generalized staggered flux phase}
\author{Tao Li}
\affiliation{ Department of Physics, Renmin University of China,
Beijing 100872, P.R.China}
\date{\today}

\begin{abstract}
We show the chiral d-wave RVB state on honeycomb lattice stands as a
natural generalization of the staggered flux phase on square
lattice. Although the state is generated from a time reversal
symmetry broken mean field ansatz, it actually represents a fully
symmetric spin liquid state with a positive definite wave function
in the sense of Marshall sign rule for unfrustrated
antiferromagnets. The evolution of the state with the parameter
$\Delta/\chi$ follows exactly the same manner as that of the
staggered flux phase on square lattice. The critical pairing
strength corresponding to the $\pi$-flux phase is found to be
$\Delta/\chi=\sqrt{2}$. As a result of the geometric frustration
between neighboring plaquette on honeycomb lattice, a direct
generalization of the $U(1)$ staggered flux pattern on square
lattice to honeycomb lattice is impossible. Replacing it is the
chiral d-wave state with $Z_{2}$ gauge structure. However, this
$Z_{2}$ gauge structure is found to be ineffective after Gutzwiller
projection and the system does not support topological degeneracy.
The chiral d-wave RVB state is also found to be a rather good
variational state for the Heisenberg model on honeycomb lattice. The
spin correlation of the chiral d-wave state is found to be greatly
enhanced as compared to the mean field prediction.
\end{abstract}
\maketitle

\section{I. Introduction}
The study of correlated electrons on honeycomb lattice has attracted
much interest recently. The honeycomb lattice has the smallest
possible coordinate number of 3 for a two dimensional lattice and
hosts Dirac-type dispersion at half filling. It is interesting to
understand how the electron correlation effect will manifest itself
in such a background.

A comparison with the square lattice case is of particular interest.
Both lattices are bipartite and the antiferromagnetic long range
order is unfrustrated in the strongly correlated limit. However, at
half filling, the honeycomb lattice has a Dirac-type dispersion and
a finite correlation strength is needed to induce antiferromagnetic
order on honeycomb lattice, while on square lattice, the system form
antiferromagnetic long range order immediately as we turn on
interaction as the half filled system has a nested Fermi surface.
More recently, it is discovered that the transition from the
semimetal phase to the antiferromagnetic ordered state on honeycomb
lattice may take place in a two step manner, leaving a small but
finite intermediate correlation range in which the system may enter
a spin liquid state with unconventional characteristics\cite{Meng}.

On square lattice, the antiferromagnetic order is destroyed by
charge carrier doping. The resultant state is believed to have
d-wave superconducting pairing and may be responsible for the
superconductivity of the high-$T_{c}$ cuprates. What is the analogy
of these doping effect for the correlated electron system on
honeycomb lattice? More specifically, what is the pairing symmetry
of the resultant superconducting state, provided that the doped
system does enter some kind of unconventional superconducting state.
This inquire is even more attractive on honeycomb lattice than on
square lattice, as the undoped honeycomb system already show
signature of spin liquid ground state, which is nothing but a
undoped version of the pairing state, the RVB state.

Superconductivity on system with a honeycomb lattice, especially
that on graphene, has been studied by many researchers from both
theoretical and experimental
perspectives\cite{Maple,Silva,Geim,Doniach,Baskaran,Neto,Hu,Baskaran1}.
However, in most of these studies, the electron correlation effect
does not play an essential role and the related pairing state has
either s-wave or extended s-wave pairing symmetry, originated from
exchanging phonon or other excitations. When the electron
correlation effect is taken into account, the chiral d-wave pairing
state\cite{Doniach,Baskaran1} has been proposed as a generalization
of the d-wave pairing state on square lattice to honeycomb
lattice(see Fig.1). However, it is not clear how the time reversal
symmetry broken chiral state are connected to the antiferromagnetic
ordered state at half filling.

At this point, it is quite useful to recall how the pairing symmetry
of the superconducting state is determined in the strong correlation
limit on square lattice. Here, a RVB state called staggered flux
phase is found to be the best variational state at half filling in
the restricted space of Fermionic RVB
state\cite{Affleck,Auerbach,Kotliar,Lee}, although it is known that
the half filled system inevitably has an antiferromagnetic long
range order. The staggered flux phase has the virtue that the
antiferromagnetic short range correlation is greatly improved by the
introduction of a non zero staggered flux. Up to a SU(2) gauge
transformation, the introduction of the staggered flux is found to
be equivalent to the introduction of d-wave pairing between the
Fermionic spinons,
 especially, when the pairing and hoping order parameter becomes identical, the staggered flux
becomes $\pi$. For general value of the staggered flux, the pairing
amplitude and the flux is related by the relation
$\tan\frac{\phi}{4}=\frac{\Delta}{\chi}$. It is thus quite natural
that the doped system will choose the d-wave pairing state as its
superconducting state as the antiferromagnetic superexchange is
already optimized by the staggered flux.

Thus, a natural questions arises as what is the analog of the
staggered flux phase on the honeycomb lattice, which can hopefully
improve the superexchange interaction by some kind of pairing(flux).
However, except for the $\pi$-flux phase, a naive generalization the
staggered flux pattern for general flux value to the honeycomb
lattice is impossible as a result of the geometric frustration
between the neighboring hexagonal plaquette(see Fig. 2). Such a
difference with square lattice originates from the frustrated nature
of the dual lattice of the honeycomb lattice, which is the
triangular lattice. Thus a generalization of the staggered flux
phase to honeycomb lattice in the space of $U(1)$ staggered flux is
impossible. However, generalization of the staggered flux phase in
the larger space of $SU(2)$ flux is still possible. In this paper,
we show that the RVB state with chiral d-wave pairing just fit such
a need.

More specifically, we show that the chiral d-wave RVB state on
honeycomb lattice can be taken as a generalized staggered flux
phase. It actually describes a spin liquid state with the full
symmetry of the Heisenberg model on honeycomb lattice and evolves in
exactly the same way as the staggered flux phase on square lattice
as a function of the parameter $\Delta/\chi$(see Fig.3). When
$\frac{\Delta}{\chi}=0$, the state reduces to the uniform RVB state
on honeycomb lattice, which has Dirac-type spinon dispersion. When
$\frac{\Delta}{\chi}=\sqrt{2}$, the state is gauge equivalent to the
$\pi$ flux phase on the honeycomb lattice. When
$\frac{\Delta}{\chi}$ exceed $\sqrt{2}$ and approaches infinite, the
state again evolves back to the uniform RVB state with zero flux. It
is interesting to note that in an earlier work on symmetric spin
liquid state on honeycomb lattice, the chiral d-wave RVB state is
classified to be a Z$_{2}$ spin liquid in the neighborhood of the
uniform RVB state\cite{Lu}. As its cousin on square lattice, the
introduction of the staggered flux improves the antiferromagnetic
correlation and we find the chiral d-wave RVB state stands as a
rather good variational state for the Heisenberg model on honeycomb
lattice. Unlike the staggered flux phase on square lattice, for
general flux value, the mean field ansatz for the chiral d-wave RVB
state posses a $Z_{2}$ rather than $U(1)$ gauge structure. We also
find that as a result of the bipartite nature of the lattice, both
the staggered flux phase on square lattice and the chiral d-wave RVB
state on honeycomb lattice satisfy the Marshall sign rule for
unfrustrated antiferromagnet. Such sign structure will prohibit the
system to show topological degeneracy in the $Z_{2}$ sense, no
matter what is the gauge structure of the mean field ansatz for the
RVB state.

This paper is organized as follows. In the next two sections, we
will review some well known results about the RVB state in general
and the staggered flux phase on square lattice in particular to set
up the stage for our discussion. In section IV, we present our
results for the chiral d-wave RVB state. Section V contains a
detailed comparison between the staggered flux phase on square
lattice and the chiral d-wave RVB state on honeycomb and a
discussion of some related issues.

\begin{figure}[h!]
\includegraphics[width=8cm,angle=0]{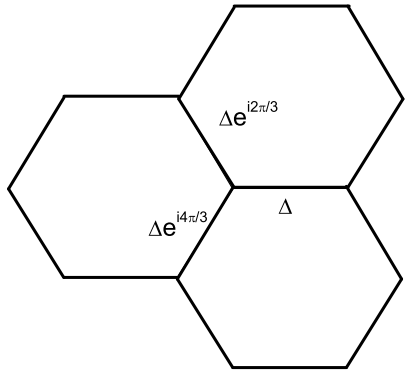}
\caption{The phases of the pairing potentials in the chiral d-wave
RVB state on honeycomb lattice. } \label{fig1}
\end{figure}

\begin{figure}[h!]
\includegraphics[width=8cm,angle=0]{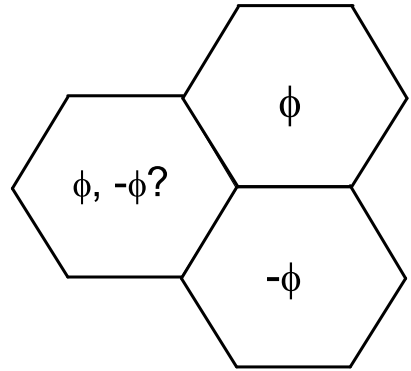}
\caption{The geometric frustration of $U(1)$ staggered flux on the
honeycomb lattice.} \label{fig2}
\end{figure}

\begin{figure}[h!]
\includegraphics[width=8cm,angle=0]{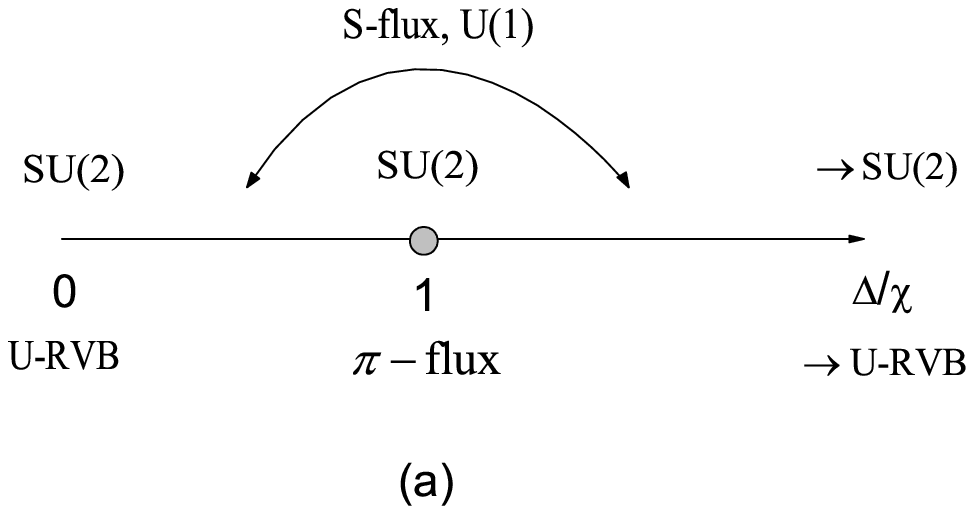}
\includegraphics[width=8cm,angle=0]{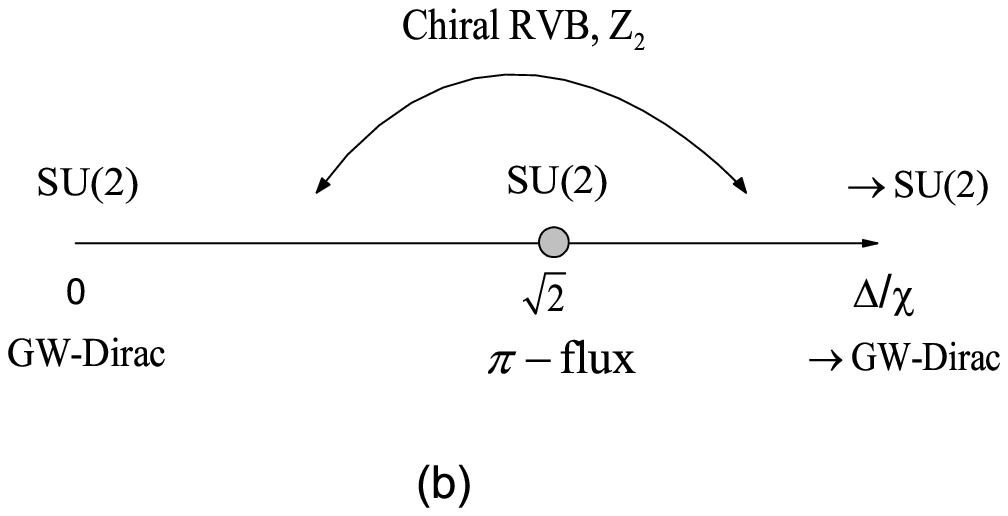}
\caption{The evolution of the gauge structure for the staggered flux
phase on square lattice(a) and the chiral d-wave RVB state on
honeycomb lattice(b). Here GW-Dirac denotes the uniform RVB state on
honeycomb lattice, which has a Dirac-type spinon dispersion.}
\label{fig3}
\end{figure}

\section{II. RVB states and their gauge structures}
The RVB states studied in this paper are those derived form
Gutzwiller projection of BCS-type mean field ground state of the
following general mean field ansatz\cite{Lee}
\begin{equation}
\mathrm{H}=-\sum_{\langle
i,j\rangle,\sigma}(\chi_{j,i}c^{\dagger}_{i,\sigma}c_{j,\sigma}+h.c.)+\sum_{\langle
i,j \rangle,\sigma}(\Delta_{i,j} \sigma
c^{\dagger}_{i,\sigma}c^{\dagger}_{j,\bar{\sigma}}+h.c.),
\end{equation}
in which $\bar{\sigma}=-\sigma$, $\chi_{i,j}=\chi^{*}_{j,i}$ and
$\Delta_{i,j}=\Delta_{j,i}$. Eq.(1) can also be taken as the saddle
point approximation of a field theoretical description of the RVB
state, in which $\chi_{i,j}$ and $\Delta_{i,j}$ are interpreted as
two types of RVB order parameters. In the field theoretical
formulation, the Gutzwiller projection into the subspace of no
double occupancy amounts to integrating over the time component of
the gauge fluctuation.

It is important to note that within the subspace of no double
occupancy, a description of the RVB state in terms of $\chi_{i,j}$
and $\Delta_{i,j}$ becomes redundant as a result of the gauge degree
of freedom in the mean field ansatz\cite{Lee}. Such redundancy is
accompanied by an underlying gauge structure of the RVB state so
constructed. To see more clearly, it is better to rewrite the mean
field ansatz in the Nambu form
\begin{equation}
\mathrm{H}=\sum_{\langle
i,j\rangle}\psi^{\dagger}_{i}U_{i,j}\psi_{j},
\end{equation}
in which $\psi^{T}_{i}=(c_{i,\uparrow},c^{\dagger}_{i,\downarrow})$
is a two component spinor. $U_{i,j}=\left( {\begin{array}{*{20}c}
  -\chi^{*}_{i,j}  & \Delta_{i,j} \\
  \Delta^{*}_{i,j} & \chi_{i,j} \\
 \end{array} } \right)$ is a $2\times 2$ matrix. It is clear then that the system is invariant under the following
 $SU(2)$ gauge transformation
\begin{eqnarray}
\psi_{i}&\rightarrow& W_{i}\psi_{i}\\ \nonumber U_{i,j}&\rightarrow&
W_{i}U_{i,j}W^{\dagger}_{j} ,
\end{eqnarray}
in which $W_{i}$ is a site-dependent $SU(2)$ matrix. As only the
$SU(2)$ gauge singlet survive the Gutzwiller projection, $U_{i,j}$
and $W_{i}U_{i,j}W^{\dagger}_{j}$ actually describes the same RVB
state.

Although $U_{i,j}$ is not gauge invariant and thus unphysical, it
does contains important gauge invariant information on the structure
of the RVB state. To uncover such internal gauge structure of the
RVB state, it is a common practice to construct the loop
operators\cite{Wen}, $P_{i}(C)=U_{i,j}U_{j,k}U_{k,l}\cdots U_{m,i}$,
in which $i$ denotes the starting point of the loop $C$ and
$j,k,\cdots m$ are the remaining sites along the loop. Under the
$SU(2)$ gauge transformation, a loop operator transform as
$P_{i}(C)\rightarrow W_{i}P_{i}(C)W^{\dagger}_{i}$ and thus its
trace form a gauge invariant quantity. A RVB state is called to have
$SU(2)$ gauge structure if all loop operator are proportional to the
identity matrix $\tau_{0}=\left( {\begin{array}{*{20}c}
 1  & 0 \\
  0 & 1 \\
 \end{array} } \right)$. Otherwise, it is called to have a $U(1)$ gauge
structure if all $P_{i}(C)$ starting from any given site $i$ commute
with each other. If there are more than two loop operators starting
from the same site do not commute with each other, the RVB state is
called to have $Z_{2}$ gauge structure. The gauge structure of the
RVB state is closely related to the existence of soft gauge mode in
the long wave length limit in the effective field theoretic
description of the RVB state. For example, if the gauge structure of
the RVB state is $SU(2)$, then there will be $SU(2)$ soft gauge mode
in the long wave length limit. While if the gauge structure of RVB
state is $Z_{2}$, then there is no soft gauge mode at all in the
long wave length limit. In such a case, the predictions made at the
level of saddle point approximation is believed to be more reliable
than those cases in which the RVB state has a $U(1)$ or $SU(2)$
gauge structure. In addition, the system with $Z_{2}$ gauge
structure is argued in the effective field theory to support
topological order and topological degeneracy - existence of
degenerate ground states that can not be differentiated from any
local probe\cite{Wen}.

The RVB state is generated from Gutzwiller projection of the
BCS-type mean field ground state
\begin{equation}
|\mathrm{RVB}\rangle=\mathrm{P}_{\mathrm{G}}|\mathrm{BCS}\rangle.
\end{equation}
Such a state can in general be rewritten in the form of condensed
spin singlet pairs
\begin{equation}
|\mathrm{RVB}\rangle=\mathrm{P}_{\mathrm{G}}\left(\sum_{i,j}a(i-j)c^{\dagger}_{i,\uparrow}c^{\dagger}_{j,\downarrow}\right)^{\frac{N}{2}}|0\rangle,
\end{equation}
in which $a(i-j)$ is the so called RVB amplitude and satisfy the
relation $a(i-j)=a(j-i)$, $N$ is the number of lattice site. As
gauge equivalent mean field ansatz $U_{i,j}$ lead to identical RVB
state, a RVB state is said to be symmetric if and only if the new
ansatz after symmetry transformation is gauge equivalent to the old
one. The projective symmetry group(PSG) provides a convenient scheme
to classify the symmetric RVB state\cite{Wen}.

For RVB state with $Z_{2}$ gauge structure, which is argued to
posses topological degeneracy from effective field theory, a $Z_{2}$
topological excitation called vison can be constructed by reversing
the sign of RVB amplitude on bonds that crossing a branch cut line
originated from the center of the vison an odd number of times. The
topological nature of the vison excitation can be seen from the fact
it acts as a $\pi$ flux tube located at its origin for spinon
traveling around it. Especially, when the RVB state is defined on a
torus and a vison is trapped in one of the holes of the torus, one
is left with a state that globally distinct from, but locally
indistinguishable(and thus degenerate in the thermodynamic limit)
from the original RVB state. This is nothing but the topological
degeneracy. However, on bipartite lattice, the topological
degeneracy predicted by the effective field theory can be lifted as
a result of some special phase structure on the RVB state.

The detection of the $Z_{2}$ topological degeneracy is simplified by
the fact that trapping a vison in the holes of the torus is gauge
equivalent to change to boundary condition from periodic to
anti-periodic or vice versa in the mean field
ansatz\cite{Ivanov,Li}. Thus to see if a particular RVB state does
posses topological degeneracy in practice, we only need to calculate
the overlap between RVB states generated from mean field ansatz with
different boundary conditions around the holes of the torus and
extrapolate the result to the thermodynamic limit. If the overlap
extrapolate to zero in the thermodynamic limit, then the RVB state
is said to posses topological degeneracy.

\section{III. The staggered flux phase on square lattice}
The staggered flux phase(or d-wave RVB state) on square lattice
plays a very important role in our understanding of the high-T$_{c}$
superconductivity in cuprates. In the following we will review
briefly some of the most important properties of this
state\cite{Lee} for a comparison with the chiral d-wave RVB state on
honeycomb lattice.

The staggered flux phase on square lattice is generated from the
following mean field ansatz
\begin{equation}
\mathrm{H}_{\mathrm{SF}}=-\sum_{\langle i,j
\rangle,\sigma}(e^{i\phi_{i,j}}c^{\dagger}_{i,\sigma}c_{j,\sigma}+h.c.),
\end{equation}
in which the phase factor $\phi_{i,j}$ is introduced to guarantee
that each plaquette of the square lattice is threaded by a $U(1)$
flux of value $\pm\Phi$ arranged in a staggered pattern. Although
the mean field ansatz breaks the time reversal symmetry for general
value of $\Phi$, it is well known that the staggered flux phase is
gauge equivalent to the d-wave RVB state generated from the
following time reversal symmetric d-wave BCS mean field ansatz
\begin{eqnarray}
\mathrm{H}_{\mathrm{d-BCS}}=&-\chi&\sum_{\langle i,j
\rangle,\sigma}(c^{\dagger}_{i,\sigma}c_{j,\sigma}+h.c.)\\\nonumber
&+&\sum_{\langle i,j
\rangle,\sigma}(\Delta_{i,j}c^{\dagger}_{i,\sigma}c^{\dagger}_{j,\bar{\sigma}}+h.c.),
\end{eqnarray}
in which $\Delta_{i,j}=\pm\Delta$ for nearest neighboring sites $i$
and $j$ in the x/y direction. Here both $\chi$ and $\Delta$ are real
and is determined from $\Phi$ by
\begin{equation}
\tan\frac{\Phi}{4}=\frac{\Delta}{\chi}.
\end{equation}
In the following, we will use the d-wave gauge.

\subsection{A. Mean field description}
The spinon excitation spectrum in the staggered flux phase has the
Dirac-type linear dispersion and is given by
$E_{\mathrm{k}}=\sqrt{(\epsilon_{\mathrm{k}})^{2}+(\Delta_{\mathrm{k}})^{2}}$,
in which
$\epsilon_{\mathrm{k}}=-2\chi(\cos\mathrm{k}_{x}+\cos\mathrm{k}_{y})$,
$\Delta_{\mathrm{k}}=2\Delta(\cos\mathrm{k}_{x}-\cos\mathrm{k}_{y})$.
The nodes are located at
$(\mathrm{k}_{x},\mathrm{k}_{y})=(\pm\pi/2,\pm\pi/2)$. As we will be
concerned with the spin structure factor of the RVB state below, we
present the mean field prediction for it here. The spin structure
factor is defined as
\begin{equation}
S(\mathrm{q})=\frac{1}{N^{2}}\sum_{i,j}e^{i\mathrm{q}\cdot(i-j)}\langle
S_{i}\cdot S_{j}\rangle.
\end{equation}
The spin structure factor in a BCS mean field ground state is given
by
\begin{equation}
S(\mathrm{q})=\frac{3}{8N^{2}}\sum_{\mathrm{k}}(1-\frac{\epsilon_{\mathrm{k}}\epsilon_{\mathrm{k+q}}+\Delta_{\mathrm{k}}\Delta_{\mathrm{k+q}}}{E_{\mathrm{k}}E_{\mathrm{k+q}}}).
\end{equation}
Since the square lattice is bipartite and
$\epsilon_{\mathrm{k}}=-\epsilon_{\mathrm{k}+\mathrm{Q}}$,
$\Delta_{\mathrm{k}}=-\Delta_{\mathrm{k}+\mathrm{Q}}$ for
$\mathrm{Q}=(\pi,\pi)$, we find $S(\mathrm{Q})=\frac{3}{4N}$ in the
mean field theory. The spin structure factor is thus independent of
the value of the staggered flux in the mean field theory. It also
extrapolates to zero in the thermodynamic limit.

\begin{figure}[h!]
\includegraphics[width=8cm,angle=0]{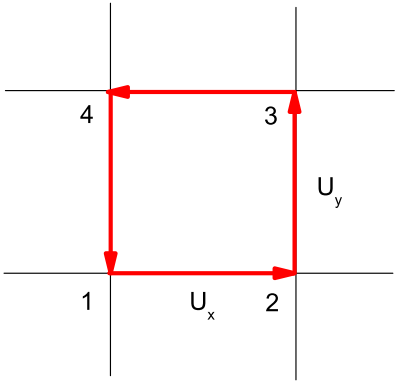}
\caption{The loop operator defined on the elementary plaquette of
square lattice.} \label{fig4}
\end{figure}
\subsection{B. Gauge structure}
The loop operator around the elementary plaquette of the square
lattice in the staggered flux phase is given by
\begin{eqnarray*}
P_{1}(C)&=&U_{x}U_{y}U_{x}U_{y}\\\nonumber
&=&(\chi^{2}-\Delta^{2})^{2}-4\chi^{2}\Delta^{2}+4i\chi\Delta(\chi^{2}-\Delta^{2})\tau_{2}\nonumber,
\end{eqnarray*}
in which $U_{x}=-\chi\tau_{3}+\Delta\tau_{1}$,
$U_{y}=-\chi\tau_{3}-\Delta\tau_{1}$. Here $\tau_{1}$, $\tau_{2}$
and $\tau_{3}$ denote the three Pauli matrixes in the internal
$SU(2)$ space.

When $\Delta/\chi=0$, $|\Delta|=|\chi|$ or $\chi/\Delta=0$, the loop
operator is proportional to $\tau_{0}$(here we assume that $\chi$
and $\Delta$ satisfy the normalization condition
$\chi^{2}+\Delta^{2}=1$). By induction we can show that all loop
operators are proportional to $\tau_{0}$ at these three points. The
state with $\chi/\Delta=0$ and that with $\Delta/\chi=0$ are in fact
gauge equivalent and is the well known uniform RVB state on square
lattice. The state with $|\Delta|=|\chi|$ is the so called
$\pi$-flux phase on square lattice. Thus, both the uniform RVB state
and the $\pi$-flux phase posses a $SU(2)$ gauge structure. For
general value of the staggered flux, the loop operator is not
proportional to $\tau_{0}$ and gauge structure can be shown to be
$U(1)$. The evolution of the gauge structure with $\Delta/\chi$ in
the staggered flux phase is summarized in Fig.3(a).

\subsection{C. Symmetries}
The staggered flux actually respects all the physical symmetries of
the Heisenberg model on square lattice. While translational symmetry
and inversion symmetry is manifest in the d-wave gauge, the
four-fold lattice rotational symmetry can be shown as follows.

Under the four-fold lattice rotation, the mean field ansatz
transform as follows: $\chi\rightarrow\chi$, $\Delta\rightarrow
-\Delta$. A global $SU(2)$ gauge transformation with
$W_{j}=i\tau_{3}$ suffices to recover the original ansatz. The
rotated ansatz and the original ansatz thus describe the same RVB
state. This prove the rotational symmetry of the staggered flux
phase.

The four-fold rotational symmetry of the staggered flux also implies
the gauge equivalence between the state with staggered flux $\Phi$
and $-\Phi$. As the gauge flux is defined modula $2\pi$, the state
with staggered flux $\Phi$ and $2\pi-\Phi$ are gauge equivalent. In
the d-wave gauge, this indicates that the state with $\Delta/\chi=a$
and the state with $\Delta/\chi=1/a$ are gauge equivalent.

\subsection{D. Sign structure of the staggered flux phase and topological degeneracy}
In an Ising basis for the spins, the staggered flux phase can be
expanded as follows
\begin{equation}
|\mathrm{SF}\rangle=\sum_{\{\sigma_{i}\}}\psi(\{\sigma_{i}\})|\{\sigma_{i}\}\rangle,
\end{equation}
in which
$|\{\sigma_{i}\}\rangle=|\sigma_{1},\cdots,\sigma_{N}\rangle$
denotes an Ising basis. Then the wave function
$\psi(\{\sigma_{i}\})$ can be shown to be real and satisfy the
Marshall sign rule\cite{Sorella,Li} for unfrustrated
antiferromagnet, namely, the sign of $\psi$ is given by
$(-1)^{N^{A\downarrow}}$, with $N^{A\downarrow}$ denoting the number
of down spins in $A$ sublattice.

Although the mean field ansatz of the staggered flux phase has at
least a $U(1)$ gauge structure, it is still possible to construct a
state with a trapped vison in the holes of a torus and check its
orthogonality with the state with no trapped vison. Such a
calculation has been done in previous studies\cite{Ivanov} and it
was found that the overlap extrapolates to a finite value in the
thermodynamic limit, indicating no topological degeneracy for the
staggered flux phase. This result is claimed to be a strong support
of the effective field theory argument. However, a later
investigation\cite{Li} indicates that the Marshall sign rule
satisfied by the staggered flux phase plays a more essential role in
removing the topological degeneracy.

\subsection{E. Variational energy and spin structure factor}
Both the short range and the long range spin correlation are seen to
reach extreme at the uniform RVB state and the $\pi$-flux phase.
This is a result of the gauge equivalence between states with
staggered flux $\Phi$ and $-\Phi$ and that between states with
staggered flux $\Phi$ and $2\pi-\Phi$, from which one can easily
show that the spin correlation function should reach their extreme
in the uniform RVB state and the $\pi$-flux phase.

The correlation between spins on nearest neighboring sites,
$\frac{1}{2N}\sum_{\langle i,j\rangle} \langle S_{i} \cdot S_{j}
\rangle$, which is proportional to the variational energy for the
Heisenberg model on square lattice, is presented in Fig.5 as a
function of $\Delta/\chi$. The optimal value for $\Delta/\chi$ is
found to be about $0.3$. The variational energy reaches local
maximum in both the uniform RVB state and the $\pi$-flux phase.

\begin{figure}[h!]
\includegraphics[width=8cm,angle=0]{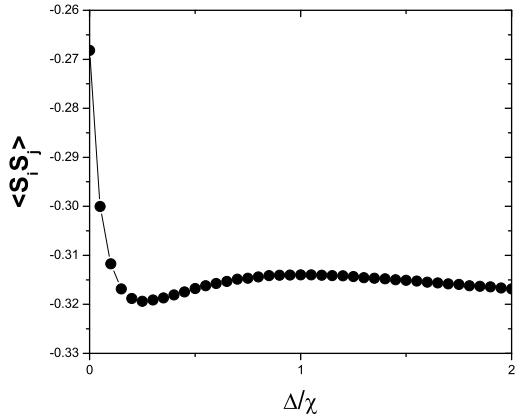}
\caption{The variational energy per bond calculated for the
Heisenberg model on square lattice as a function of $\Delta/\chi$.
The calculation is done on a $16\times 16$ lattice with
periodic-antiperiodic boundary condition. The error bars are smaller
than the size of the symbols.} \label{fig5}
\end{figure}

To probe the long range spin correlation, we have calculated the
spin structure factor of the staggered flux phase. The spin
structure factor peaks at the antiferromagnetic ordering wave vector
$Q=(\pi,\pi)$. In Fig.6, we present the result for $S(Q)$ as a
function of $\Delta/\chi$. The spin structure factor decrease
monotonically with the staggered flux for $\Phi<\pi$. The small peak
close to the uniform RVB state is a finite size effect and it moves
toward the uniform RVB state with increasing lattice size and
disappears in the thermodynamic limit.

In a recent work\cite{Li1}, we have shown that the uniform RVB state
on square lattice actually describe a state with antiferromagnetic
long range order as a result of the nested spinon Fermi surface. The
spin structure factor follows the $S(Q)\approx
S_{0}+\alpha(1/L)^{5/4}$ behavior as a function of the linear scale
of the lattice, $L$. While in the $\pi$-flux phase, $S(Q)$ is found
to follow the $S(Q)\approx (1/L)^{3/2}$ behavior and no magnetic
order is detected. In between the two extremes, the staggered flux
has no magnetic long range order and the spin correlation function
decay algebraically with distance with an exponent depending on the
value of the staggered flux. Thus the uniform RVB state state on
square lattice is an isolated singular point.
\begin{figure}[h!]
\includegraphics[width=8cm,angle=0]{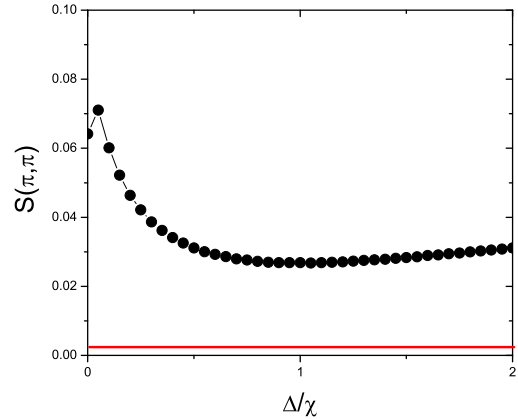}
\caption{The spin structure factor at $\mathrm{q}=(\pi,\pi)$ as a
function of $\Delta/\chi$ for the staggered flux phase on square
lattice. The calculation is done on a $16\times 16$ lattice with
periodic-antiperiodic boundary condition. The error bars are smaller
than the size of the symbols.The red line denotes the mean field
prediction, which is independent of $\Delta/\chi$.} \label{fig6}
\end{figure}

\begin{figure}[h!]
\includegraphics[width=8cm,angle=0]{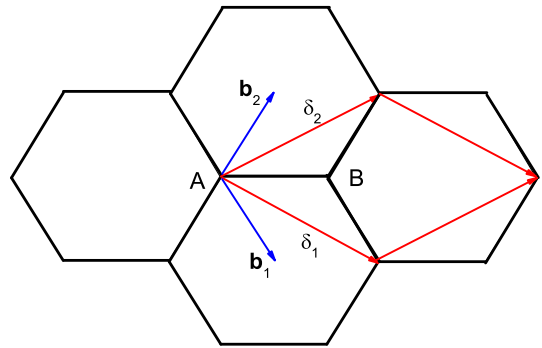}
\caption{The unit cell of the honeycomb lattice. } \label{fig7}
\end{figure}

\section{IV. The chiral d-wave RVB state on honeycomb lattice}
The chiral d-wave RVB state on honeycomb lattice is generated from
the following BCS mean field ansatz
\begin{equation}
\mathrm{H}_{\mathrm{BCS}}=-t\sum_{\langle
i,j\rangle,\sigma}(c^{\dagger}_{i,\sigma}c_{j,\sigma}+h.c.)+\sum_{\langle
i,j \rangle,\sigma}\Delta_{i,j}( \sigma
c^{\dagger}_{i,\sigma}c^{\dagger}_{j,\bar{\sigma}}+h.c.),
\end{equation}
in which $\sum_{\langle i,j \rangle}$ denotes sum over nearest
neighboring sites on honeycomb lattice. $\Delta_{i,j}$ denotes the
pairing order parameter between site $i$ and $j$. In the chiral
d-wave RVB state, $\Delta_{i,j}$ is a complex number and the phase
of $\Delta_{i,j}$ in the three directions differ with each other by
$\pm\frac{2\pi}{3}$(see Fig.1).

\subsection{A. Mean field description}
The mean field Hamiltonian for the chiral d-wave state can be easily
diagonalized in momentum space in which it takes the form
\begin{equation}
\mathrm{H}_{\mathrm{BCS}}=\sum_{\mathrm{k}}\psi^{\dagger}_{\mathrm{k}}\mathrm{M}\psi_{\mathrm{k}},
\end{equation}
in which $\psi_{\mathrm{k}}$ is a four component spinor given by
$\psi_{\mathrm{k}}^{T}=(c^{A}_{\mathrm{k}\uparrow},c^{B}_{\mathrm{k}\uparrow},c^{A\dagger}_{\mathrm{-k}\downarrow},c^{B\dagger}_{\mathrm{-k}\downarrow})$
. Here $A$ and $B$ are sublattice index. The $4\times4$ matrix
$\mathrm{M}$ is given by
\begin{equation}
\mathrm{M}=\left( {\begin{array}{*{20}c}
  0 & \gamma_{\mathrm{k}} & 0 & \Delta_{\mathrm{k}} \\
  \gamma^{*}_{\mathrm{k}} & 0 & \Delta_{\mathrm{-k}} & 0 \\
  0 & \Delta^{*}_{\mathrm{-k}} & 0 & -\gamma_{\mathrm{k}} \\
  \Delta^{*}_{\mathrm{k}} & 0 & -\gamma^{*}_{\mathrm{k}} & 0 \\
 \end{array} } \right) \nonumber,
\end{equation}
in which
$\gamma_{\mathrm{k}}=(1+e^{-i\mathrm{k}\cdot\delta_{1}}+e^{-i\mathrm{k}\cdot\delta_{2}})$,
$\Delta_{\mathrm{k}}=\Delta(1+e^{i\frac{2\pi}{3}}e^{-i\mathrm{k}\cdot\delta_{1}}+e^{i\frac{4\pi}{3}}e^{-i\mathrm{k}\cdot\delta_{2}})$.
Here $\delta_{1}$ and $\delta_{2}$ are the two lattice translational
vectors of the honeycomb lattice(see Fig.7).

The eigenvalues of the Hamiltonian are given by
\begin{equation}
\mathrm{E}^{2}_{\mathrm{k}\pm}=\gamma^{2}_{\mathrm{k}}+
\frac{|\Delta_{\mathrm{k}}|^{2}+|\Delta_{\mathrm{-k}}|^{2}\pm\sqrt{(|\Delta_{\mathrm{k}}|^{2}-|\Delta_{\mathrm{-k}}|^{2})^{2}+4|g_{\mathrm{k}}|^{2}}}{2},
\end{equation}
in which
$g_{\mathrm{k}}=\gamma_{-\mathrm{k}}\Delta^{*}_{-\mathrm{k}}-\gamma_{\mathrm{k}}\Delta^{*}_{\mathrm{k}}$.

When $\Delta=0$, the dispersion reduces to that of the Dirac
Fermion, $\mathrm{E}_{\mathrm{k}}=\pm|\gamma_{\mathrm{k}}|$. At
$(k_{1},k_{2})=\pm(2\pi/3,-2\pi/3)$, $\mathrm{E}_{\mathrm{k}}=0$.
Here we have used the convention that
$\mathrm{k}=k_{1}\mathrm{\mathbf{b}}_{1}+k_{2}\mathrm{\mathbf{b}}_{2}$
for the momentum. $\mathrm{\mathbf{b}}_{1}$ and
$\mathrm{\mathbf{b}}_{2}$ are the two reciprocal vectors. They are
dual to the lattice translational vectors $\delta_{1}$ and
$\delta_{2}$ and satisfy the relation $\delta_{i}\cdot
\mathrm{\mathbf{b}}_{j}=\delta_{i,j}$.

When $(k_{1},k_{2})=\pm(2\pi/3,-2\pi/3)$, $\gamma_{\mathrm{k}}=0$
and thus $g(\mathrm{k})=0$. It can also be shown that one of the two
gap functions, $\Delta_{\mathrm{k}}$ or $\Delta_{-\mathrm{k}}$, is
zero at $(k_{1},k_{2})=\pm(2\pi/3,-2\pi/3)$. It is then easy to
verify that the nodes at $(k_{1},k_{2})=\pm(2\pi/3,-2\pi/3)$ persist
for any value of $\Delta$. In addition, new nodes may emerge at
nonzero $\Delta$. In the appendix, we show the dispersion for some
typical values of $\Delta/\chi$. It should be noted that the
location of the additional nodes move continuously in the momentum
space with $\Delta/\chi$.

On the honeycomb lattice, the spin structure factor at the
antiferromagnetic ordering wave vector can be defined as follows
\begin{equation}
S(Q)=\frac{1}{N^{2}}\sum_{i,j}\langle(S^{A}_{i}-S^{B}_{i})\cdot(S^{A}_{j}-S^{B}_{j})\rangle.
\end{equation}
From the mean field theory, it is straightforward (though somewhat
tedious) to show that $S(Q)$ is independent of $\Delta/\chi$ and is
given by the $S(Q)=\frac{3}{4N}$. As for the staggered flux phase on
square lattice, the $\Delta/\chi$ independence of $S(Q)$ on
honeycomb lattice at the mean field level is a result of the
bipartite nature of the lattice.

\begin{figure}[h!]
\includegraphics[width=8cm,angle=0]{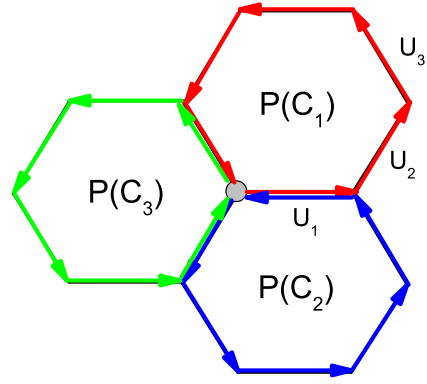}
\caption{The loop operators of the chiral d-wave RVB state on the
elementary plaquette of the honeycomb lattice. The gray dots denotes
the common starting point of the loop operators.} \label{fig8}
\end{figure}

\subsection{B. Gauge structure}
From every sites of the honeycomb lattice, there are three
elementary loop operators. They are given by
\begin{eqnarray}
P(C_{1})&=&(U_{1}U_{2}U_{3})^{2}\\\nonumber
P(C_{2})&=&(U_{2}U_{3}U_{1})^{2}\\\nonumber
P(C_{3})&=&(U_{3}U_{1}U_{2})^{2},
\end{eqnarray}
in which $U_{1}=-\chi\tau_{3}+\Delta\tau_{1}=\rho u_{1}$,
$U_{2}=-\chi\tau_{3}-\frac{\Delta}{2}\tau_{1}+\frac{\sqrt{3}\Delta}{2}\tau_{2}=\rho
u_{2}$,
$U_{3}=-\chi\tau_{3}-\frac{\Delta}{2}\tau_{1}-\frac{\sqrt{3}\Delta}{2}\tau_{2}=\rho
u_{3}$ are RVB order parameters on bonds in three directions.
$\rho=\sqrt{\chi^{2}+\Delta^{2}}$ is the 'length' of the RVB order
parameter. $u_{1},u_{2},u_{3}$ are three Hermitian matrix satisfying
$u_{1}u_{1}=u_{2}u_{2}=u_{3}u_{3}=\tau_{0}$. The products of
$U_{1}$, $U_{2}$ and $U_{3}$ in Eq.17 are given by
\begin{eqnarray}
U_{1}U_{2}U_{3}&=&-i\frac{3\sqrt{3}}{2}\chi\Delta^{2}\\\nonumber
&+&(\chi^{2}-\frac{\Delta^{2}}{2})(-\chi\tau_{3}+\Delta\tau_{1}-\sqrt{3}\Delta\tau_{2})\\\nonumber
U_{2}U_{3}U_{1}&=&-i\frac{3\sqrt{3}}{2}\chi\Delta^{2}\\\nonumber
&+&(\chi^{2}-\frac{\Delta^{2}}{2})(-\chi\tau_{3}+\Delta\tau_{1}+\sqrt{3}\Delta\tau_{2})\\\nonumber
U_{3}U_{1}U_{2}&=&-i\frac{3\sqrt{3}}{2}\chi\Delta^{2}\\\nonumber
&+&(\chi^{2}-\frac{\Delta^{2}}{2})(-\chi\tau_{3}-2\Delta\tau_{1}).
\end{eqnarray}

From Eq.18, it can be easily seen that at the three special points
$\Delta=0$, $\Delta=\sqrt{2}\chi$ and $\chi=0$(with the
normalization condition $\Delta^{2}+\chi^{2}=1$ assumed), all the
three loop operators are proportional to the identity matrix
$\tau_{0}$. By induction it can then be shown that all loop
operators are proportional to $\tau_{0}$. The system thus has
$SU(2)$ gauge structure at these three special points. This is
exactly what happens in the staggered flux phase on square lattice.
In fact, it can further shown that the chiral RVB state on honeycomb
lattice with $\Delta=0$ ($\chi=0$) and $\Delta=\sqrt{2}\chi$ are
just the uniform RVB state and $\pi$-flux phase on honeycomb
lattice.

The gauge equivalence between the state with $\Delta=0$ and $\chi=1$
($U_{1}=U_{2}=U_{3}=-\tau_{3}$) and that with $\chi=0$ and
$\Delta=1$ ($U_{1}=\tau_{1}$,
$U_{2}=-\frac{1}{2}\tau_{1}+\frac{\sqrt{3}}{2}\tau_{2}$,
$U_{3}=-\frac{1}{2}\tau_{1}-\frac{\sqrt{3}}{2}\tau_{2}$) can be
shown by the following $SU(2)$ gauge transformation. First, the
phase of the pairing order parameter in the state with $\chi=0$ and
$\Delta=1$ can be gauged away by a $U(1)$ gauge transformation.
After this $U(1)$ gauge transformation, the mean field ansatz takes
the form $U_{1}=U_{2}=U_{3}=\tau_{1}$. Then a uniform $SU(2)$
rotation with $W_{i}=e^{-i\frac{\pi}{4}\tau_{2}}$ suffices to
transform the ansatz into the form $U_{1}=U_{2}=U_{3}=-\tau_{3}$.
Similarly, the gauge equivalence between the state with
$\Delta=\sqrt{2}\chi$ and the $\pi$-flux phase on honeycomb
lattice(see Fig.9) can be established with the $SU(2)$ gauge
transformation given in Table.1 .
\begin{table}
  \centering
  \begin{tabular}{|c|c|c|c|c|}
    \hline
    Sublattice/Parity & (o,o) & (o,e) & (e,o) & (e,e) \\
    \hline
    A & $\tau_{0}$ & $-iu_{3}$ & $-iu_{2}$ & $-iu_{1}$ \\
    \hline
    B & $u_{1}\tau_{3}$ & $-u_{2}\tau_{3}$ & $-u_{3}\tau_{3}$ & $i\tau_{3}$ \\
    \hline
  \end{tabular}
  \caption{The gauge transformation that relate the chiral d-wave RVB state with $\Delta/\chi=\sqrt{2}$ and the $\pi$-flux phase shown in Fig.9}\label{1}
\end{table}

\begin{figure}[h!]
\includegraphics[width=8cm,angle=0]{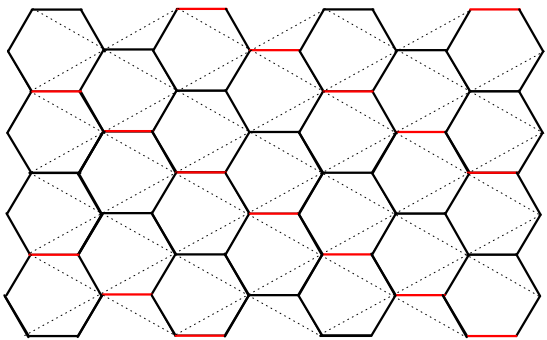}
\caption{The mean field ansatz for the $\pi$-flux on honeycomb
lattice. The bonds in red have a negative hopping integral, while
other bonds have a positive hopping integral.} \label{fig8}
\end{figure}

For other value of $\Delta/\chi$, the three loop operators do not
commute with each other and the system has a $Z_{2}$ gauge
structure. This is quite different from the square lattice where the
$SU(2)$ points are connected by intermediate state with $U(1)$ gauge
structure. The origin of such a difference can be traced back to the
geometric frustration in the dual lattice of honeycomb lattice. To
see the evolution of the gauge structure more clearly, we rewrite
the loop operator in the following form
\begin{eqnarray}
P(C_{1})&=&-\rho^{6}e^{i\Phi\vec{n}_{1}\cdot \vec{\tau}}\\\nonumber
P(C_{2})&=&-\rho^{6}e^{i\Phi\vec{n}_{2}\cdot \vec{\tau}}\\\nonumber
P(C_{3})&=&-\rho^{6}e^{i\Phi\vec{n}_{3}\cdot \vec{\tau}},
\end{eqnarray}
in which the absolute value of the gauge flux $\Phi$ is determined
by
\begin{equation}
\cos\frac{\Phi}{2}=\frac{3\sqrt{3}\chi\Delta^{2}}{2\rho^{3}}.
\end{equation}
$\vec{n}_{1}$, $\vec{n}_{2}$ and $\vec{n}_{3}$ are three vectors of
unit length and are given by
\begin{eqnarray}
\vec{n}_{1}&=&r(\Delta,-\sqrt{3}\Delta,-\chi)\\\nonumber
\vec{n}_{2}&=&r(\Delta,\sqrt{3}\Delta,-\chi)\\\nonumber
\vec{n}_{3}&=&r(-2\Delta,0,-\chi),
\end{eqnarray}
in which $r=1/\sqrt{\chi^{2}+4\Delta^{2}}$. The three vectors form
the equal angle with each other as shown schematically in Fig.10.
When $\Delta=0$, all the three vectors lies in the $-\tau_{3}$
direction. With increasing $\Delta/\chi$, the three vectors split
from each other. When $\Delta/\chi\rightarrow \infty$,
$\vec{n}_{1}$, $\vec{n}_{2}$ and $\vec{n}_{3}$ evolves into the
$\tau_{1}-\tau_{2}$ plane with a $120^{0}$ angle between each other.

\begin{figure}[h!]
\includegraphics[width=8cm,angle=0]{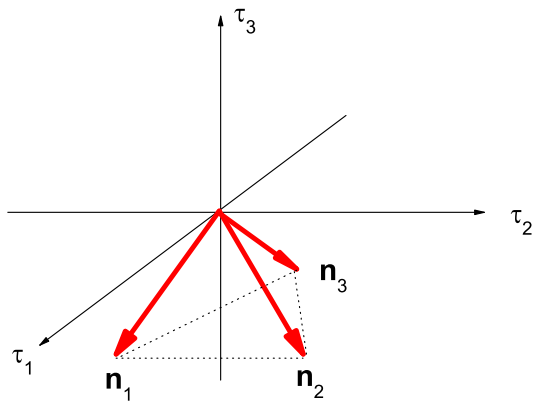}
\caption{The direction of the three loop operators in the internal
$SU(2)$ gauge space.} \label{fig9}
\end{figure}

The structure of the loop operators presented above can be viewed as
a generalized staggered flux pattern on honeycomb lattice. The
absolute value of the flux for the staggered flux phase on square
lattice and the chiral d-wave RVB state on honeycomb lattice are
both uniform. It is the direction of the flux that is responsible
for its staggered character. On square lattice, where the dual
lattice is also square lattice, a two sublattice (antiferromagnetic)
arrangement of the direction of the flux is the most natural choice.
While on honeycomb lattice, where the dual lattice is triangular
lattice, a three sublattice arrangement of the direction for the
flux is more natural.

The discussion above on the gauge structure of the chiral d-wave RVB
state is summarized in Fig.3b. Its similarity with the figure for
the staggered flux phase on square lattice is apparent. It is for
this reason that we call the chiral d-wave RVB state a generalized
staggered flux phase on honeycomb lattice.

\subsection{C. Symmetries}
As will be clear below, the term 'chiral' is in fact not quite
accurate. It can be shown that the time reversal symmetry breaking
manifested in the mean field ansazt is in fact an artifact of the
mean field description. In fact, it can be more generally proved
that the chiral d-wave RVB state actually respects all the physical
symmetries for the Heisenberg model on honeycomb lattice. The effect
of the symmetry transformation on the mean field ansatz is to induce
permutations between $U_{1}$, $U_{2}$ and $U_{3}$. Thus to prove the
symmetry of the chiral d-wave RVB state it suffice to show that the
RVB order parameters after such permutation are gauge equivalent to
the original one.

For example, under the time reversal transformation, $U_{2}$ and
$U_{3}$ are exchanged. Such a change can also be induced by the
following two-step gauge transformation. First, we rotate the ansatz
uniformly along the $\tau_{2}$ axis by $\pi$. Then as the lattice is
bipartite, we effect a $U(1)$ gauge transformation by $\pi$
uniformly on sites in the B sublattice. This proves the time
reversal symmetry of the chiral d-wave RVB state. Note that the
bipartite nature of the lattice is essential for the restoration of
the time reversal symmetry. Another example is provided by the
$2\pi/3$ rotation of the lattice, which induces a cyclic exchange
among $U_{1}$, $U_{2}$ and $U_{3}$. It can be easily shown that such
a cyclic exchange can also be induced by a uniform rotation of the
ansatz along the $\tau_{3}$-axis in the $SU(2)$ space by an angle
$2\pi/3$. This proves the rotational symmetry of the chiral d-wave
RVB state.

We note that as the exchange between $U_{2}$ and $U_{3}$ and the
cyclic exchange among $U_{1}$,$U_{2}$ and $U_{3}$ exhausts the
generators of the permutation group of $U_{1}$,$U_{2}$ and $U_{3}$,
we have in fact proved the full symmetry of the chiral d-wave RVB
state.

\begin{figure}[h!]
\includegraphics[width=8cm,angle=0]{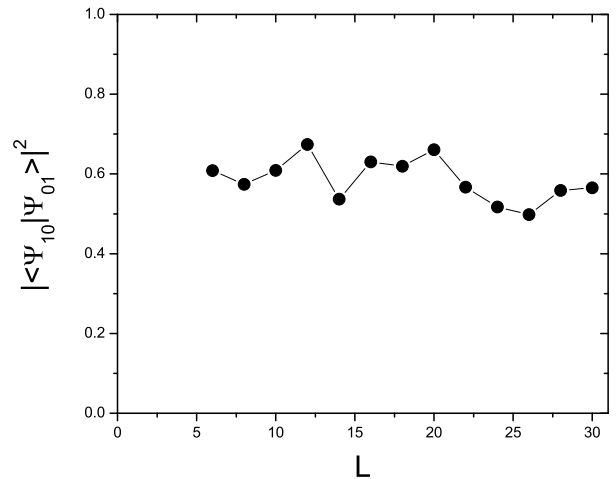}
\caption{The overlap between $|\Psi_{10}\rangle$ and
$|\Psi_{01}\rangle$ on a $L \times L \times2$ honeycomb lattice as a
function of the lattice size $L$. The error bars are smaller than
the size of the symbols. The oscillation of the curve is caused by
the complex nodal structure of the chiral d-wave RVB state, which
move continuously in the momentum space with $\Delta/\chi$.}
\label{fig11}
\end{figure}

\subsection{D. Sign structure of the chiral d-wave RVB state and topological degeneracy}
Beside being time reversal symmetric, it can also be shown that the
chiral d-wave RVB state on honeycomb lattice actually has a positive
definite wave function in the sense of Marshall sign rule for
bipartite antiferromagnet, although the mean field ground state
break the time reversal symmetry and has a complex valued wave
function. This remarkable result is in fact a general property for
all RVB states generated from bipartite mean field ansatz\cite{Li}.
It is also argued that the Marshall sign structure will remove from
the projected wave function the topological degeneracy, even if the
mean field ansatz has a $Z_{2}$ gauge structure.

The absence of the topological degeneracy on bipartite system can be
argued as follows. As changing the boundary condition around the
hole of a tours will not change the bipartite nature of the mean
field ansatz, both the projected state with trapped vison and
without trapped vison will be positive definite in the sense of
Marshall sign rule. It is just this insensitivity of the sign
structure to the trapped $Z_{2}$ gauge flux that is responsible for
the absence of the topological degeneracy. In a recent
work\cite{Li2}, we have shown through VMC that the chiral d-wave RVB
state on honeycomb lattice does not support topological degeneracy.
However, since the calculation is done in the region
$\Delta/\chi<\sqrt{2}$, we would like to supplement it with the
result in the region $\Delta/\chi>\sqrt{2}$.

The overlap between states with different number of vison in both
holes of the torus for the chiral d-wave RVB state is shown in
Fig.11. Here $|\Psi_{10}\rangle$ denotes the state with trapped
vison in the hole surrounded by the x-circumference but no vison in
the hole surrounded by the y-circumference. $|\Psi_{01}\rangle$)
denotes the state with trapped vison in the hole surrounded by the
y-circumference but no vison in the hole surrounded by the
x-circumference. The calculation is done at $\Delta/\chi=5$. The
overlap is seen to oscillate with the lattice size. Such oscillation
is caused by the complex nodal structure of the chiral d-wave RVB
state, which move continuously in the momentum space with
$\Delta/\chi$. However, it is clear that the overlap will
extrapolate to finite value in the thermodynamic limit.

\begin{figure}[h!]
\includegraphics[width=8cm,angle=0]{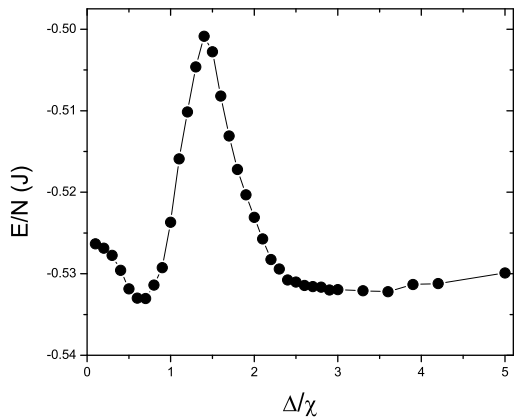}
\caption{The variational energy per site for the Heisenberg model on
honeycomb lattice calculated from the chiral d-wave RVB state. The
calculation is done on a $16\times 16\times 2$ lattice with
periodic-antiperiodic boundary condition. The error bars are smaller
than the size of the symbols.} \label{fig12}
\end{figure}

\begin{figure}[h!]
\includegraphics[width=8cm,angle=0]{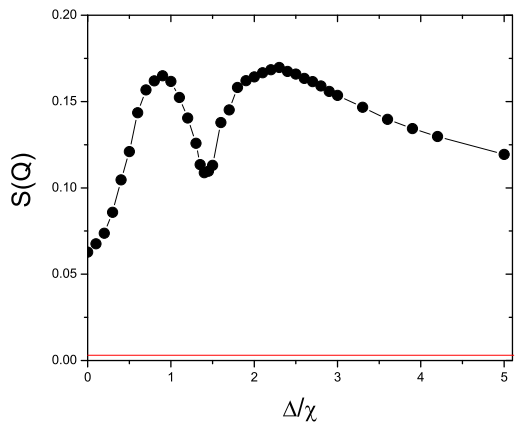}
\caption{The spin structure factor at the antiferromagnetic ordering
wave vector for the chiral d-wave RVB state on honeycomb lattice as
a function $\Delta/\chi$. The calculation is done on a $16\times
16\times 2$ lattice with periodic-antiperiodic boundary condition.
The error bars are smaller than the size of the symbols. The small
non-analycity of the curve is checked to be finite size effect
induced by the moving nodes in the momentum space. The red line
denotes the mean field prediction, which is independent of
$\Delta/\chi$.} \label{fig13}
\end{figure}

\subsection{E. Variational energy and spin structure factor}
The variational energy for the Heisenberg model on honeycomb lattice
is shown in Fig.12 as a function of $\Delta/\chi$. As for the
staggered flux phase on square lattice, the variational energy is
optimized with a state intermediate between uniform RVB state and
the $\pi$-flux phase. Unlike the square lattice case, the difference
in energy between the uniform RVB state and the optimized
variational state in this class of RVB state is seen to be small. As
the exact ground state energy is estimated to be $-0.55J$ per site
from exact diagonalizition studies\cite{Jafari}, we see the chiral
d-wave RVB state in fact represents a rather good variational state
for the Heisenberg model on honeycomb lattice.

The spin structure factor at the antiferromagnetic ordering wave
vector for the chiral d-wave RVB state is plotted in Fig.13 as a
function of $\Delta/\chi$. As a comparison we also plot the value
calculated from the mean field ground state($=\frac{3}{4N}$), which
is independent of the pairing strength. The spin structure factor is
seen to be reduced at the uniform RVB state and the $\pi$ flux
phase, both of which having a $SU(2)$ gauge structure at the mean
field level. This can be taken as a evidence that the $SU(2)$ gauge
fluctuation will reduce rather than enhance the spin correlation.

We have also calculated the size dependence of the spin structure
factor at $\Delta/\chi=0$, $\Delta/\chi=1$ and
$\Delta/\chi=\sqrt{2}$ to see how the gauge structure affects the
spin correlation in the thermodynamic limit. The results are shown
in Fig.14. We find in all of the three states the spin structure
factor decay algebraically with the linear scale of the lattice,
indicating no magnetic long range order in the thermodynamic limit.
This is different from square lattice, where the uniform RVB state
is found to posses antiferromagnetic long range order. A fit to the
date with the formula $S(Q)=\alpha(1/L)^{\beta}$ shows that the
exponent $\beta$ varies with the value of $\Delta/\chi$. More
specifically, the exponent at $\Delta/\chi=0$ is found to be
approximately $1.5$. It is reduced to $1$ at $\Delta/\chi=1$ and
return to $1.3$ at $\Delta/\chi=\sqrt{2}$. Thus, a finite value for
the flux will not only improve the local spin correlation, but also
result in more long ranged spin correlation.

\begin{figure}[h!]
\includegraphics[width=8cm,angle=0]{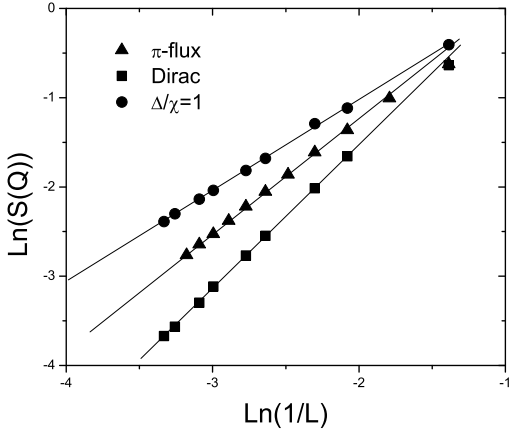}
\caption{The finite size scaling behavior of the spin structure
factor of the chiral d-wave RVB state for $\Delta/\chi=0$(Dirac),
$\Delta/\chi=1$ and $\Delta/\chi=\sqrt{2}$($\pi$-flux).}
\label{fig14}
\end{figure}

\section{V. Conclusion}
From the above discussion, we see the chiral d-wave RVB state on
honeycomb lattice indeed stands as a natural generalization of the
staggered flux phase on square lattice. The two states shares the
following properties.

(1) Both states respect the full symmetry of the Heisenberg model on
respective lattices and have a positive definite wave function in
the sense of the Marshall sign rule.

(2) Both states evolves with $\Delta/\chi$ in the same manner. With
the increase of $\Delta/\chi$, both states evolve from the uniform
RVB state to the $\pi$-flux phase and then back to the uniform RVB
state. The uniform RVB state and the $\pi$-flux phase on both
lattices have a promoted gauge symmetry of $SU(2)$.

(3)The introduction of the gauge flux on both lattice improves the
short range spin correlation and both the staggered flux phase on
square lattice and the chiral d-wave RVB state on honeycomb lattice
are rather good variational description of the Heisenberg model on
respective lattices.

(4)As a result of the bipartite nature of the lattice and the
resultant Marshall sign rule of the RVB state, both the staggered
flux phase on square lattice and the chiral d-wave RVB state on
honeycomb lattice do not support topological degeneracy, although
the latter posses a $Z_{2}$ gauge structure at the saddle point
level.

However, there are also important differences between the staggered
flux phase on square lattice and the chiral d-wave RVB state on
honeycomb lattice. The following is a list of the main differences.

(1)On square lattice, the intermediate state connecting the uniform
RVB state and the $\pi$-flux phase has a $U(1)$ gauge structure.
However, on honeycomb lattice, the two states are connected by
intermediate state with $Z_{2}$ gauge structure. This difference
originates from the frustrated nature of the dual lattice of
honeycomb lattice. Related with this difference is the difference in
the staggering pattern of the gauge flux on the two lattices. On
square lattice, the gauge flux form a two sublattice collinear
pattern. On the other hand, the gauge flux on the honeycomb lattice
form a three sublattice non-collinear pattern.

(2)As a result of the non-collinear nature of the gauge flux in the
chiral d-wave state, the states with gauge flux $\Phi$ below
$\Delta/\chi=\sqrt{2}$ are not gauge equivalent to the states with
gauge flux $2\pi-\Phi$ above $\Delta/\chi=\sqrt{2}$, except the
special case of the uniform RVB state. Thus in principle the
$\pi$-flux phase on honeycomb lattice need not be a extreme of
physical properties as functions of $\Delta/\chi$, although we find
both the variational energy and the spin structure factor do reach
their extreme at the $\pi$-flux phase. On the other hand, on square
lattice, the $\pi$-flux phase must be an extreme of physical
properties as functions of $\Delta/\chi$ as protected by the
$\Phi\rightarrow2\pi-\Phi$ symmetry.

(3)On square lattice, the uniform RVB state posses antiferromagnetic
long range order. However, on honeycomb lattice, no magnetic order
is detected in the uniform RVB state. This difference can be induced
by the difference in their spinon dispersion. While the uniform RVB
state on square lattice has a large and nested spinon Fermi surface,
which is unstable toward antiferromagnetic ordering, the uniform RVB
state on honeycomb lattice has an isotropic Dirac-type spinon
dispersion, just as the $\pi$-flux phase on square lattice. Indeed,
the spin structure factor of the uniform RVB state on honeycomb
lattice is found to show the same scaling behavior as the $\pi$-flux
phase on square lattice\cite{Xu}.

(4)Although the variational energy can be improved by introducing
gauge flux on both square lattice and honeycomb lattice, the amount
of improvement is quite different. On square lattice, the
introduction of staggered flux will lead to a $15\%$ improvement in
the variational energy of the Heisenberg model. On the other hand,
the same energy gain on honeycomb lattice is less than $2\%$. This
difference can also be induced by the difference in the spinon
dispersion on the two lattices.

As the staggered flux phase plays such a central role in our
understanding of the high-Tc superconductivity in cuprates from the
perspective of RVB theory, it is quite natural to expect that the
chiral d-wave RVB state will play the similar role, if the doped
honeycomb system in the strongly correlated regime does support some
kind of superconductivity. After doping, the time reversal symmetry
will be broken and the state will become truly chiral. A full gap
will also be opened in the doped system, as can be shown by
including a chemical potential term in Eq.(13). There will be many
interesting issues concerning this novel superconducting state,
especially on its topological properties. However, as the
condensation energy is much smaller than system on square lattice,
the superconductivity on the honeycomb system should be weaker than
high T$_{c}$ cuprates. It is interesting to see if this state can be
realized in experiment.

This work is supported by NSFC Grant No. 10774187 and National Basic
Research Program of China No. 2007CB925001 and No. 2010CB923004. The
author is grateful for the discussions with Fan Yang and Cenke Xu.

\section{Appendix}

In this appendix, we present the mean field spinon dispersion of the
chiral d-wave RVB state on honeycomb lattice for some typical values
of $\Delta/\chi$. The result is shown in Fig.15.
\begin{figure}[h!]
\includegraphics[width=9cm,angle=0]{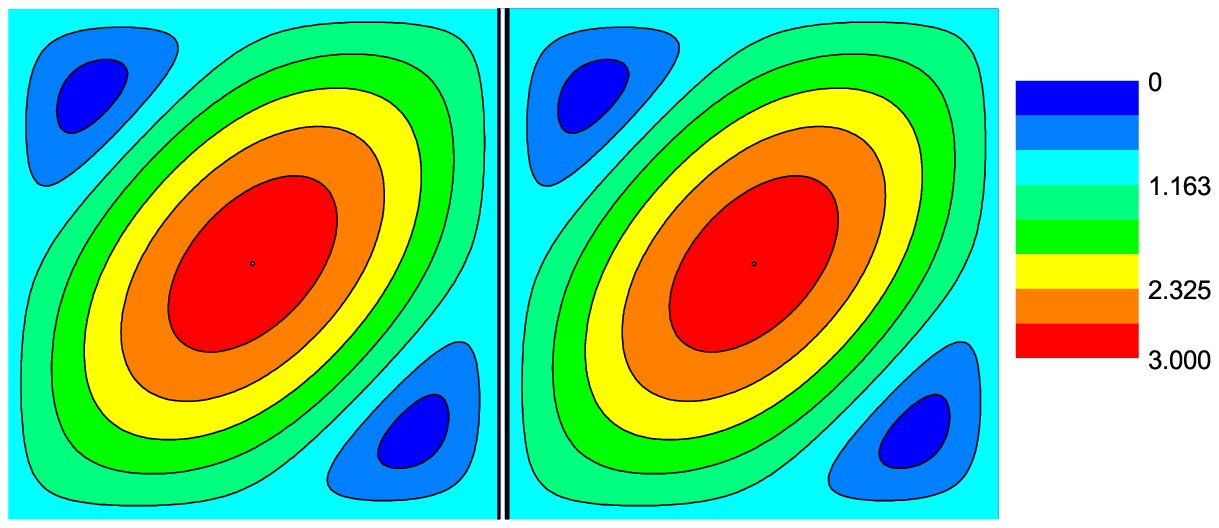}
\includegraphics[width=9cm,angle=0]{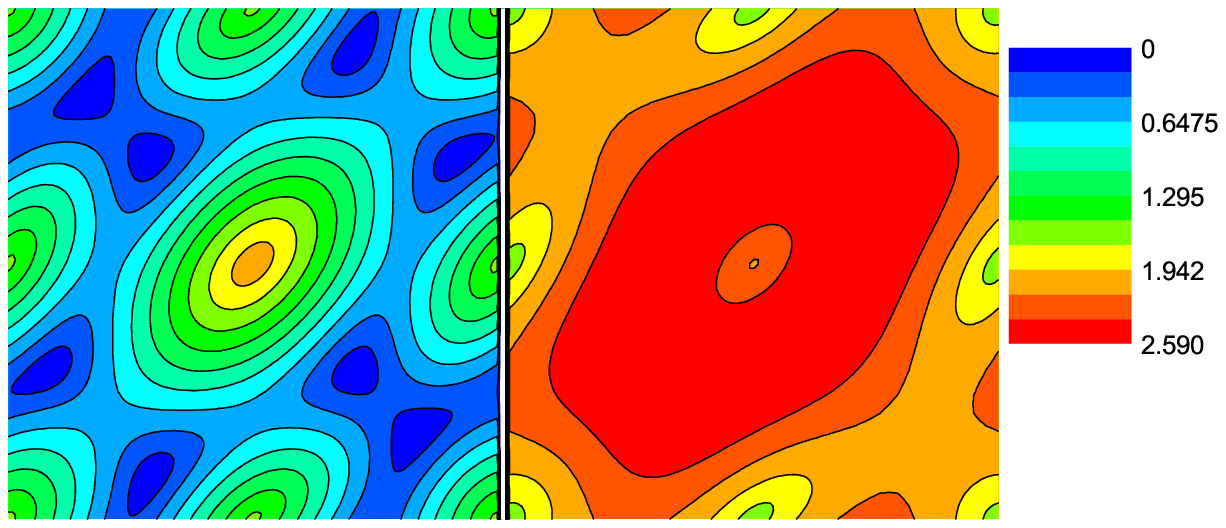}
\includegraphics[width=9cm,angle=0]{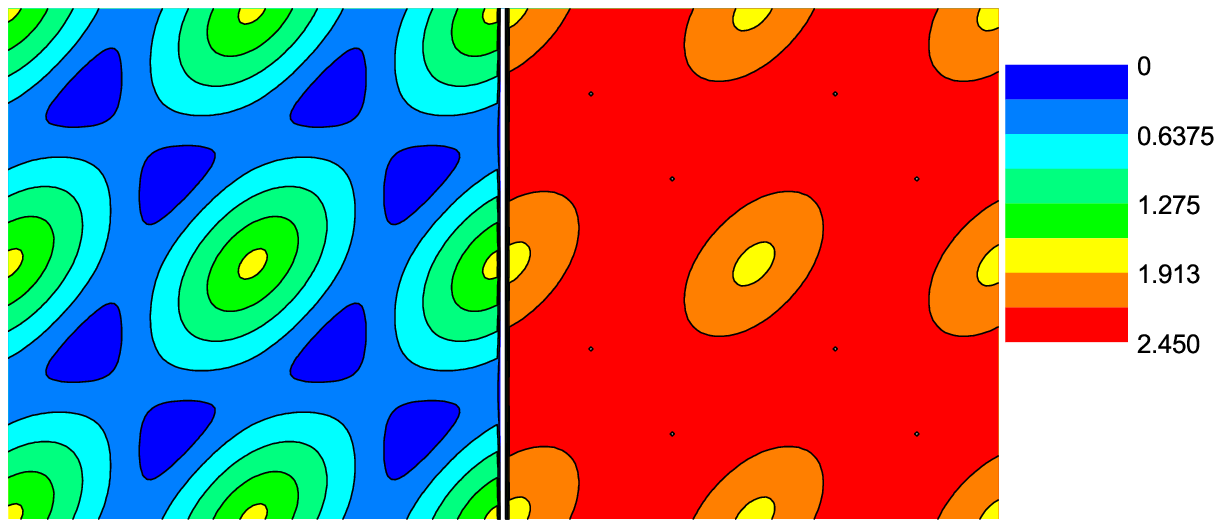}
\includegraphics[width=9cm,angle=0]{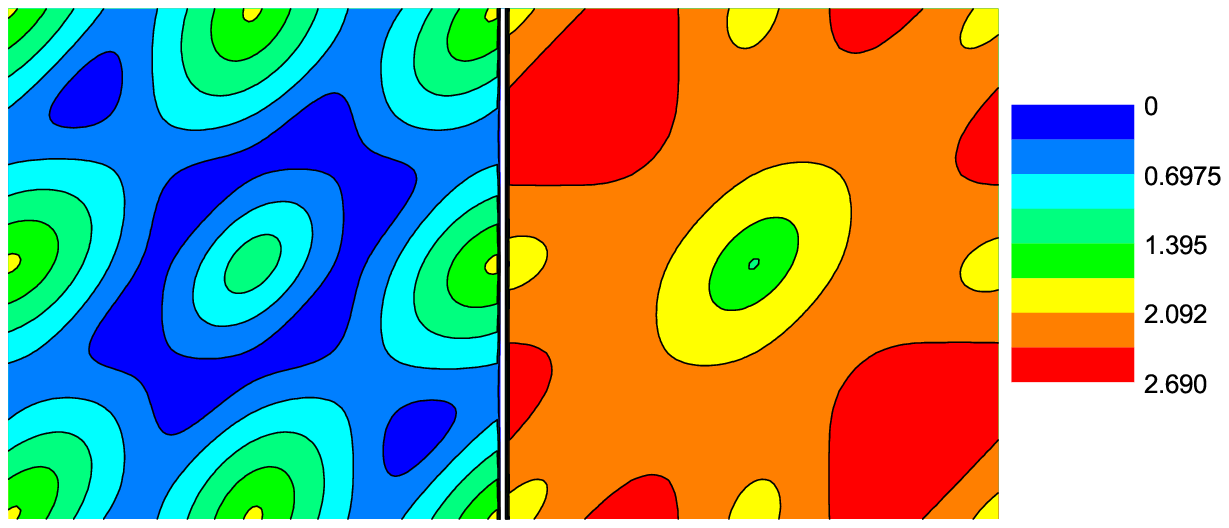}
\includegraphics[width=9cm,angle=0]{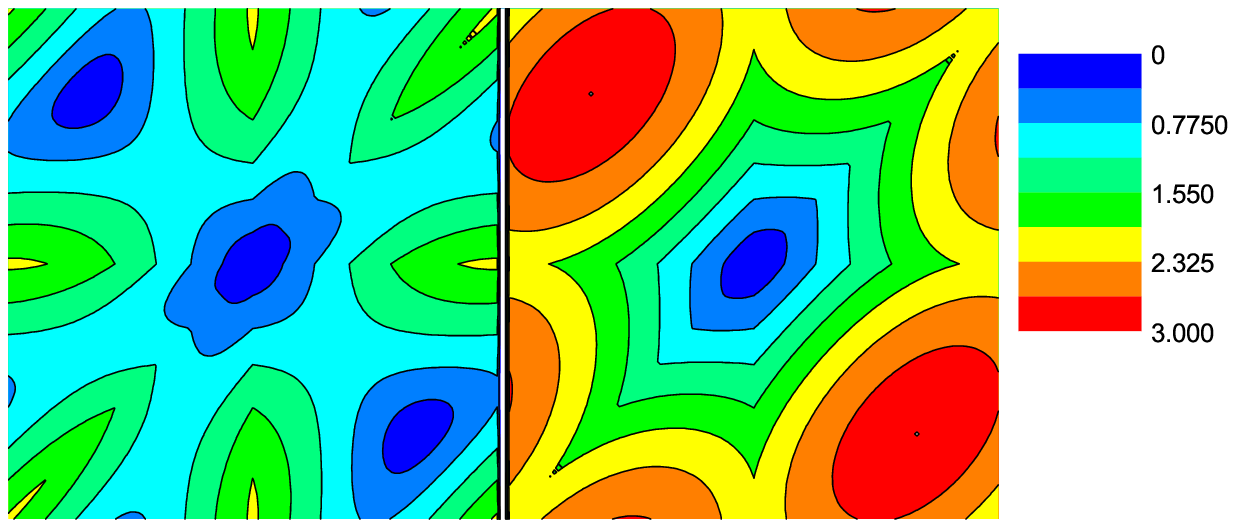}
\caption{The evolution of the mean field spinon dispersion as a
function of $\Delta/\chi$. From top to bottom are the results for
$\Delta/\chi=0$, $\Delta/\chi=1$, $\Delta/\chi=\sqrt{2}$,
$\Delta/\chi=2$ and $\chi/\Delta=0$. The value of $\chi$ and
$\Delta$ are normalized so that $\chi^{2}+\Delta^{2}=1$. The lower
branches($\mathrm{E}^{-}_{\mathrm{k}}$) are plotted in the left
panel and the upper branch ($\mathrm{E}^{+}_{\mathrm{k}}$) are
plotted in the right panel.} \label{fig15}
\end{figure}

\end{document}